# Multivariate Cryptography-Based Anonymous Certificate Scheme

Abel C. H. Chen
*Information & Communications Security Laboratory,*
*Chunghwa Telecom Laboratories*
Taoyuan, Taiwan
ORCID: 0000-0003-3628-3033

***Abstract*—As quantum computing technology continues to mature, the US National Institute of Standards and Technology (NIST) has outlined a migration timeline for Post-Quantum Cryptography (PQC), recommending the deprecation of certain elliptic curve cryptography (ECC) by 2030. Furthermore, privacy-sensitive application scenarios, such as vehicular communications, require the use of anonymous certificates. However, existing anonymous certificate schemes are still largely based on ECC. Therefore, this study proposes a multivariate cryptography-based anonymous certificate scheme, aiming to design quantum-safe anonymous certificates suitable for privacy-sensitive application services. The proposed multivariate cryptography-based anonymous certificate scheme is supported by rigorous mathematical proofs and illustrated with computational cases.**

## I. INTRODUCTION

In recent years, continuous technological breakthroughs in the field of quantum computing [1] have accelerated the anticipated arrival of Q-Day by several years [2], thereby posing a potential threat to the security of RSA and elliptic curve cryptography (ECC) [3]. Consequently, in November 2024, the US National Institute of Standards and Technology (NIST) published the draft of NIST IR 8547, recommending a transition to Post-Quantum Cryptography (PQC). The draft further outlines a plan to deprecate certain RSA and ECC schemes by 2030 and to fully prohibit the use of RSA and ECC by 2035 [4]. These developments indicate that the migration of Public Key Infrastructure (PKI) to PQC has become an urgent and ongoing imperative.

To establish PQC standards, NIST has conducted multiple rounds of PQC standardization conferences and evaluations, resulting in the selection of several candidate algorithms [5], [6]. The currently dominant approaches include lattice-based cryptography, for which the Module-Lattice-Based Key-Encapsulation Mechanism (ML-KEM) for encryption and key encapsulation [7], the Module-Lattice-Based Digital Signature Algorithm (ML-DSA) for digital signatures [8], and the Falcon algorithm, which is still under standardization [5], have been proposed. Furthermore, the Stateless Hash-Based Digital Signature Algorithm (SLH-DSA) based on hash-based cryptography [9], and the Hamming Quasi-Cyclic (HQC) key-encapsulation mechanism based on code-based cryptography [6], have also been selected by NIST for standardization. Nevertheless, the aforementioned digital signature schemes may suffer from relatively large signature sizes. To address this limitation, NIST has initiated an additional digital signature scheme standardization process, aiming to identify solutions with shorter signature lengths and higher signing efficiency. Within this process, several multivariate cryptography-based digital signature algorithms (DSAs), such as the Unbalanced Oil and Vinegar (UOV) scheme, have been selected as candidate proposals [10].

Furthermore, certain application scenarios (e.g., healthcare systems [11] and vehicular communications [12], [13]) have stringent requirements for privacy protection, which motivate the development of cryptographic mechanisms such as anonymous certificates. In the context of vehicular communications, the IEEE 1609.2 standard [12] and the IEEE 1609.2.1 standard [13] introduce the Butterfly Key Expansion (BKE) method. By deriving butterfly public keys that cannot be used to infer the original caterpillar public keys, BKE enables the deployment of anonymous certificates and provides privacy protection. However, the butterfly key expansion mechanisms specified in IEEE 1609.2 and IEEE 1609.2.1 are built upon ECC and therefore may be vulnerable to quantum computing attacks.

Therefore, the primary objective of this study is to design an anonymous certificate scheme that is resistant to quantum computing attacks. Furthermore, reports from the additional digital signature scheme standardization process indicate that multivariate cryptography offers advantages such as short signature lengths and high signing efficiency. Accordingly, this study proposes a multivariate cryptography-based anonymous certificate scheme. The main contributions of this work are summarized as follows.

- This study proposes a multivariate cryptography-based key expansion method built upon the Matsumoto-Imai cryptosystem.
- Based on the proposed key expansion method, this study further proposes a multivariate cryptography-based anonymous certificate scheme that provides privacy protection.
- This study presents theoretical analysis, security proofs, and illustrative computational cases, and describes the implementation methods and tools used in the experimental environment.

This paper is organized into five sections. Section II introduces the core principles of the Matsumoto-Imai cryptosystem. Section III presents the proposed multivariate cryptography-based key expansion method and the proposed multivariate cryptography-based anonymous certificate scheme. Section IV describes the implementations and computational cases. Finally, Section V summarizes the contributions of this study and discusses future work.

## II. MATSUMOTO-IMAI CRYPTOSYSTEM

This section presents the key pair generation, encryption/decryption, and digital signature based on the Matsumoto-Imai cryptosystem.

### A. Key Pair Generation

Let $r$ be a finite field of characteristic two, namely $r = GF(2)$, with cardinality $c$. Let $g(x) \in r[x]$ be an irreducible polynomial of degree $n$. The field $R = r[x]/g(x)$ is then defined as a degree-$n$ extension field of $r$.

A mapping $h$ is defined to associate each element of $R$ with its coefficient vector in $r^n$, that is, $h: R \to r^n$. The inverse mapping $h^{-1}: r^n \to R$ reconstructs the corresponding field element from its coefficient vector. Accordingly, for any element $a_0 + a_1x^1 + \cdots + a_{n-1}x^{n-1} \in R$, the mapping $h$ is defined as $h(\sum_{i=0}^{n-1} a_ix^i) = (a_0, a_1, \ldots, a_{n-1})$, and the inverse mapping $h^{-1}$ is defined as $h^{-1}(a_0, a_1, \ldots, a_{n-1}) = (a_0 + a_1x^1 + \cdots + a_{n-1}x^{n-1}) = \sum_{i=0}^{n-1} a_ix^i$.

Let $\tau$ be an integer satisfying $0 < \tau < n$ and $gcd(c^\tau + 1, c^n - 1) = 1$. When $\tau$ is chosen such that this condition holds, a mapping $K$ can be defined over the field $R$. Owing to the above condition on $\tau$, this mapping is invertible. The mapping $K$ is defined as $K(X) = X^{c^\tau+1}$. Assume that there exists an integer $t$ satisfying $t(c^\tau + 1) \equiv 1 \ (mod\ c^n - 1)$. Under this assumption, the inverse mapping $K^{-1}$ can be expressed as $K^{-1}(X) = X^t$. Proofs of the fact that $K$ and $K^{-1}$ are mutual inverses can be found in [14] and [15]. The mappings $K$ and $K^{-1}$ serve as the core public and private functions, respectively, in the Matsumoto-Imai cryptosystem.

Let $\widetilde{K}$ be a mapping defined over $r^n$ as $\widetilde{K}(a_0, a_1, \ldots, a_{n-1}) = h \circ K \circ h^{-1}(a_0, a_1, \ldots, a_{n-1}) = (\tilde{k}_0, \tilde{k}_1, \ldots, \tilde{k}_{n-1})$, where $\tilde{k}_0, \tilde{k}_1, \ldots, \tilde{k}_{n-1} \in r[x]$. Let $M_1$ and $M_2$ be two invertible affine transformations defined on $r^n$. The public mapping $\overline{K}$ is then defined as $\overline{K}(a_0, a_1, \ldots, a_{n-1}) = M_1 \circ \widetilde{K} \circ M_2(a_0, a_1, \ldots, a_{n-1}) = (\bar{k}_0, \bar{k}_1, \ldots, \bar{k}_{n-1})$, where $\bar{k}_0, \bar{k}_1, \ldots, \bar{k}_{n-1} \in r[x]$. As the public key, the mapping $\overline{K}$ can be used by any party for encryption or signature verification. The affine transformations $M_1$ and $M_2$ are primarily employed to conceal the structure of the core function $K$. This design principle is analogous to that of the McEliece cryptosystem [16] and aims to reduce the cryptanalysis problem to an NP-hard problem in terms of computational complexity [14], [16].

Accordingly, the public key $\overline{K}(a_0, a_1, \ldots, a_{n-1})$ is defined over the finite field $r$, while the private key $\overline{K}^{-1}(b_0, b_1, \ldots, b_{n-1})$ consists of the inverse function $K^{-1}$ together with the inverses of the two affine transformations, ${M_1}^{-1}$ and ${M_2}^{-1}$. Explicitly, the private inversion is given by ${M_2}^{-1} \circ \widetilde{K}^{-1} \circ {M_1}^{-1}(b_0, b_1, \ldots, b_{n-1}) = {M_2}^{-1} \circ h^{-1} \circ K^{-1} \circ h \circ {M_1}^{-1}(b_0, b_1, \ldots, b_{n-1})$.

### B. Encryption and Decryption

In encryption and decryption applications, it is assumed that the plaintext has been encoded as a set of coefficient vectors $A = \{a_0, a_1, \ldots, a_{n-1}\}$. The coefficient vector is then substituted into the public key $\overline{K}$ to perform encryption, and the resulting coefficient vector set is obtained as the ciphertext $B = \{b_0, b_1, \ldots, b_{n-1}\}$, as shown in Eq. (1).

$$\begin{aligned} \overline{K}(A) &= \overline{K}(a_0, a_1, \ldots, a_{n-1}), \\ &= (b_0, b_1, \ldots, b_{n-1}) = B. \end{aligned} \tag{1}$$

During decryption, the coefficient vector of the ciphertext is substituted into the private key $\overline{K}^{-1}$. The resulting coefficient vector set corresponds to the recovered plaintext $\{a_0, a_1, \ldots, a_{n-1}\}$, as shown in Eq. (2).

$$\begin{aligned} \overline{K}^{-1}(B) &= \overline{K}^{-1}(b_0, b_1, \ldots, b_{n-1}), \\ &= (a_0, a_1, \ldots, a_{n-1}) = A. \end{aligned} \tag{2}$$

The mathematical proof of the correctness of the encryption and decryption processes is given in Eq. (3), as detailed in [14] and [15].

$$\begin{aligned} \overline{K}^{-1}(B) &= \overline{K}^{-1}(\overline{K}(A)), \\ &= {M_2}^{-1} \circ \widetilde{K}^{-1} \circ {M_1}^{-1}\left(M_1 \circ \widetilde{K} \circ M_2(A)\right), \\ &= {M_2}^{-1} \circ h^{-1} \circ K^{-1} \circ h\left(h \circ K \circ h^{-1} \circ M_2(A)\right), \\ &= {M_2}^{-1}(M_2(A)), \\ &= A = (a_0, a_1, \ldots, a_{n-1}). \end{aligned} \tag{3}$$

### C. Digital Signature

In digital signature applications, it is assumed that the message has been encoded as a set of coefficient vectors $P = \{p_0, p_1, \ldots, p_{n-1}\}$. The coefficient vector is then substituted into the private key $\overline{K}^{-1}$ to generate the signature, and the resulting coefficient vector set is obtained as the signature value $S = \{s_0, s_1, \ldots, s_{n-1}\}$, as shown in Eq. (4).

$$\begin{aligned} \overline{K}^{-1}(P) &= \overline{K}^{-1}(p_0, p_1, \ldots, p_{n-1}), \\ &= (s_0, s_1, \ldots, s_{n-1}) = S. \end{aligned} \tag{4}$$

For signature verification, the coefficient vector of the signature value is substituted into the public key $\overline{K}$. The resulting coefficient vector set is obtained as the verification value $V = \{v_0, v_1, \ldots, v_{n-1}\}$, as shown in Eq. (5). If the verification value $V$ is identical to the message $P$, the signature is considered valid; otherwise, the verification fails.

$$\begin{aligned} \overline{K}(S) &= \overline{K}(s_0, s_1, \ldots, s_{n-1}), \\ &= (v_0, v_1, \ldots, v_{n-1}) = V. \end{aligned} \tag{5}$$

The mathematical proof of the correctness of the signature generation and verification processes is given in Eq. (6), as detailed in [14] and [15].

$$\begin{aligned} V &= \overline{K}(S) = \overline{K}(\overline{K}^{-1}(P)), \\ &= M_1 \circ \widetilde{K} \circ M_2\left({M_2}^{-1} \circ \widetilde{K}^{-1} \circ {M_1}^{-1}(P)\right), \\ &= M_1 \circ h \circ K \circ h^{-1}\left(h^{-1} \circ K^{-1} \circ h \circ {M_1}^{-1}(P)\right), \\ &= M_1\left({M_1}^{-1}(P)\right) \\ &= P = (p_0, p_1, \ldots, p_{n-1}). \end{aligned} \tag{6}$$

## III. THE PROPOSED METHODS AND SCHEMES

This study first presents, in Section III.A, a key expansion method designed based on the Matsumoto-Imai cryptosystem, such that the expanded public key and the expanded private key remain computationally paired. Section III.B introduces the proposed anonymous certificate scheme, and Section III.C further integrates the butterfly key expansion concept to achieve anonymity even with respect to both the Certificate Authority (CA) and Registration Authority (RA).

### A. Key Expansion Method

To achieve anonymity and to prevent the derivation of any linkage between the original public key $\overline{K}$ and the expanded public key $\overline{K}'$, a key expansion method is designed in this section. Specifically, a random seed $e$ is used to generate a random invertible matrix (i.e., an invertible affine transformation) $M_{e,j}$. The original public key is then composed with this random invertible matrix to obtain the expanded public key, as shown in Eq. (7). Correspondingly, the inverse of the random invertible matrix, ${M_{e,j}}^{-1}$, is composed with the original private key to derive the expanded private key $\bar{K}'^{-1}$, as shown in Eq. (8).

$$\begin{aligned}\bar{K}'(a_0, a_1, \dots, a_{n-1}) &= M_{e,j} \circ \bar{K}(a_0, a_1, \dots, a_{n-1}), \\ &= M_{e,j} \circ M_1 \circ \tilde{K} \circ M_2(a_0, a_1, \dots, a_{n-1}).\end{aligned} \tag{7}$$

$$\begin{aligned}&\bar{K}'^{-1}(b_0, b_1, \dots, b_{n-1}) \\ &= \bar{K}^{-1} \circ {M_{e,j}}^{-1}(b_0, b_1, \dots, b_{n-1}), \\ &= {M_2}^{-1} \circ \tilde{K}^{-1} \circ {M_1}^{-1} \circ {M_{e,j}}^{-1}(b_0, b_1, \dots, b_{n-1}).\end{aligned} \tag{8}$$

Using encryption and decryption as an example, Eq. (9) demonstrates that the expanded public key $\bar{K}'$ and the expanded private key $\bar{K}'^{-1}$ remain a valid key pair. By the same reasoning, it can be shown that, in digital signature applications, the expanded public key $\bar{K}'$ and the expanded private key $\bar{K}'^{-1}$ also form a valid pair. For efficient generation of the random invertible matrix (i.e., invertible affine transformation) $M_{e,j}$, previously published work by the author [16] can be referenced, which describes a method to generate a random invertible matrix and its inverse with time complexity O($n$).

$$\begin{aligned}\bar{K}'^{-1}(B) &= \bar{K}'^{-1}(\bar{K}'(A)), \\ &= \bar{K}^{-1} \circ {M_{e,j}}^{-1}\left(M_{e,j} \circ \bar{K}(A)\right) = \bar{K}^{-1}(\bar{K}(A)), \\ &= {M_2}^{-1} \circ \tilde{K}^{-1} \circ {M_1}^{-1}\left(M_1 \circ \tilde{K} \circ M_2(A)\right), \\ &= {M_2}^{-1} \circ h^{-1} \circ K^{-1} \circ h\left(h \circ K \circ h^{-1} \circ M_2(A)\right), \\ &= {M_2}^{-1}(M_2(A)), \\ &= A = (a_0, a_1, \dots, a_{n-1}).\end{aligned} \tag{9}$$

By applying the random invertible matrix $M_{e,j}$, the original public key is protected, preventing an adversary from deriving the original public key $\bar{K}$ from the expanded public key $\bar{K}'$. As a result, anonymity and privacy protection are achieved.

### *B. The Proposed Anonymous Certificate Scheme*

This section assumes a setting in which an anonymous certificate is issued by a CA to an End Entity (EE). The anonymous certificate scheme proposed in this study is primarily built upon the key expansion method proposed in Section III.A, as illustrated in Fig. 1.

First, the EE generates an original key pair, consisting of the original public key $\bar{K}$ and the original private key $\{{M_1}^{-1}, {M_2}^{-1}, K^{-1}\}$. The EE then encrypts ($\bar{K}$, $I$) using the public key of the CA to produce a ciphertext $w_1$, which is subsequently transmitted to the CA. Here, $I$ denotes the content of to-be-signed-certificate, such as the application permissions of the EE.

Upon receiving the ciphertext $w_1$, the CA decrypts the ciphertext $w_1$ using its private key to obtain ($\bar{K}$, $I$). The CA then generates a random value $e_2$ and applies a secure pseudorandom function (e.g., a keyed-Hash Message Authentication Code (HMAC)) by using $e_2$ as the HMAC key and an integer $j$ as the HMAC message, in order to derive pseudorandom values and an invertible matrix (i.e., an invertible affine transformation) $M_{e_2,j}$. The detailed construction of this process can be found in [16]. The CA subsequently composes the original public key $\bar{K}$ with the random invertible matrix $M_{e,j}$ to obtain the expanded public key $\bar{K}' = M_{e_2,j} \circ \bar{K}$. The certificate authority then generates an anonymous certificate ***Cert*** for the end entity, in which the public key field contains the expanded public key $\bar{K}'$. Finally, the CA encrypts (***Cert***, $e_2$) using the original public key of the EE to obtain a ciphertext $w_3$, and transmits $w_3$ to the EE.

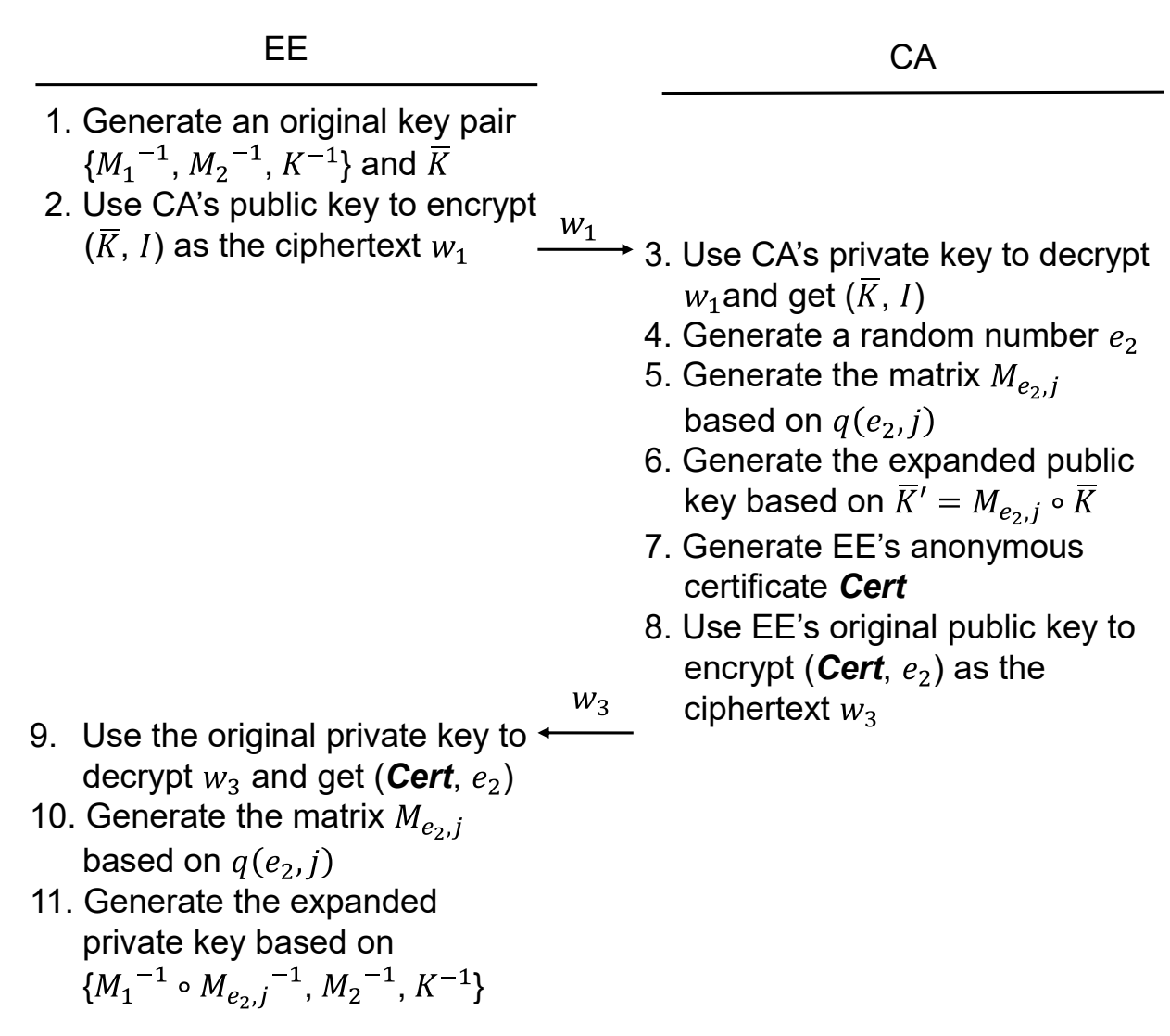

Fig. 1. The proposed anonymous certificate scheme.

Upon receiving the ciphertext $w_3$, the EE decrypts the ciphertext $w_3$ using its original private key to obtain (***Cert***, $e_2$). The EE then retrieves the random value $e_2$ and uses it as the HMAC key, with the integer $j$ as the HMAC message, to derive pseudorandom values and the inverse matrix ${M_{e_2,j}}^{-1}$. Based on this result, the expanded private key $\{{M_1}^{-1} \circ {M_{e_2,j}}^{-1}, {M_2}^{-1}, K^{-1}\}$ is computed.

Because the public key field of the anonymous certificate contains the expanded public key $\bar{K}'$, and because the original public key $\bar{K}$ cannot be derived from $\bar{K}'$ without knowledge of the random value $e_2$ or the invertible matrix $M_{e_2,j}$, the proposed anonymous certificate scheme satisfies the anonymity requirement. However, under this scheme, the CA is still able to determine the relationship between the expanded public key and the original public key. To further achieve anonymity with respect to the CA, an anonymous certificate scheme based on butterfly key expansion is designed in Section III.C.

### *C. The Proposed Anonymous Certificate Scheme Based on Butterfly Key Expansion*

By performing one key expansion at the RA and a second key expansion at the CA, followed by the issuance of an anonymous certificate to the EE, anonymity with respect to both the RA and the CA is achieved, as illustrated in Fig. 2.

First, the EE generates a caterpillar key pair, consisting of the caterpillar public key $\bar{K}$ and the caterpillar private key $\{{M_1}^{-1}, {M_2}^{-1}, K^{-1}\}$, and generates a random value $e_1$. The EE then encrypts ($\bar{K}$, $e_1$, $I$) using the public key of the RA to produce a ciphertext $w_1$, which is transmitted to the RA. Here, $I$ denotes the content of to-be-signed-certificate, such as the application permissions of the EE.

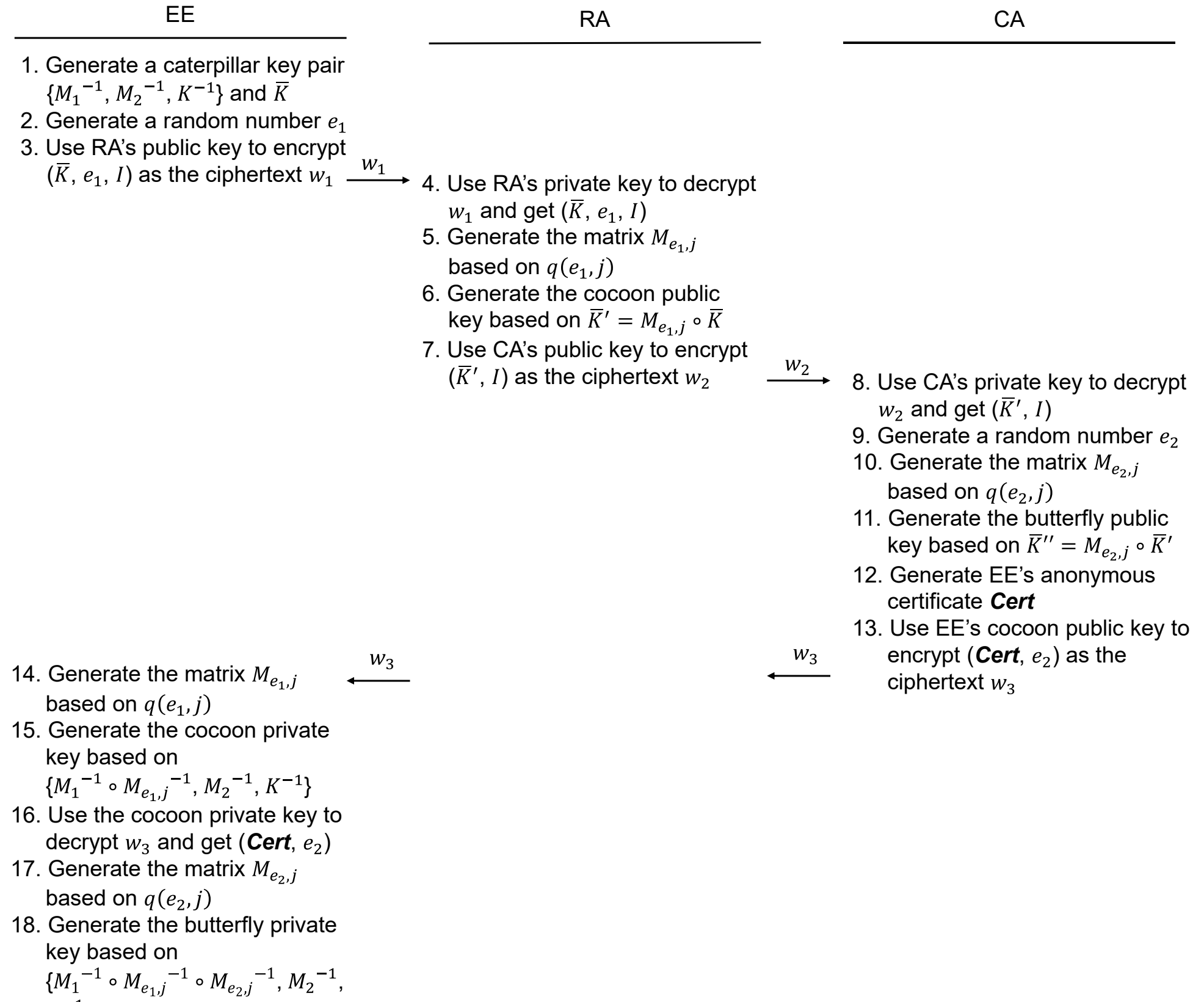

Fig. 2. The proposed anonymous certificate scheme based on butterfly key expansion.

Upon receiving the ciphertext $w_1$, the RA decrypts the ciphertext $w_1$ using its private key to obtain $(\overline{K}, e_1, I)$. The registration authority then retrieves the random value $e_1$ and applies a secure pseudorandom function (e.g., HMAC) using $e_1$ as the HMAC key and an integer $j$ as the HMAC message, to generate pseudorandom values and an invertible matrix (i.e., an invertible affine transformation) $M_{e_1,j}$. The registration authority subsequently composes the caterpillar public key $\overline{K}$ with the random invertible matrix $M_{e_1,j}$ to obtain the cocoon public key $\overline{K}' = M_{e_1,j} \circ \overline{K}$. The registration authority then encrypts $(\overline{K}', I)$ using the public key of the certificate authority to produce a ciphertext $w_2$, and transmits $w_2$ to the CA.

Upon receiving the ciphertext $w_2$, the CA decrypts the ciphertext $w_2$ using its private key to obtain $(\overline{K}', I)$. The CA then generates a random value $e_2$ and applies a secure pseudorandom function (e.g., HMAC) using $e_2$ as the HMAC key and an integer $j$ as the HMAC message, to derive pseudorandom values and an invertible matrix (i.e., an invertible affine transformation) $M_{e_2,j}$. The CA subsequently composes the cocoon public key $\overline{K}'$ with the random invertible matrix $M_{e_2,j}$ to obtain the butterfly public key $\overline{K}'' = M_{e_2,j} \circ \overline{K}'$. The CA then generates an anonymous certificate ***Cert*** for the EE, in which the public key field contains the butterfly public key $\overline{K}''$. Finally, the certificate authority encrypts (***Cert***, $e_2$) using the cocoon public key of the EE to produce a ciphertext $w_3$, which is transmitted to the EE.

The EE uses the random value $e_1$ as the HMAC key and the integer $j$ as the HMAC message to derive pseudorandom values and the inverse matrix $M_{e_1,j}{}^{-1}$, and thereby computes the cocoon private key $\{M_1^{-1} \circ M_{e_1,j}{}^{-1}, M_2^{-1}, K^{-1}\}$. Upon receiving the ciphertext $w_3$, the EE decrypts the ciphertext $w_3$ using the cocoon private key to obtain (***Cert***, $e_2$). The EE then retrieves the random value $e_2$ and applies it as the HMAC key, with the integer $j$ as the HMAC message, to derive pseudorandom values and the inverse matrix $M_{e_2,j}{}^{-1}$. Based on this result, the butterfly private key is computed as $\{M_1^{-1} \circ M_{e_1,j}{}^{-1} \circ M_{e_2,j}{}^{-1}, M_2^{-1}, K^{-1}\}$.

## D. Proofs

Using encryption and decryption as an example, Eq. (10) demonstrates that the butterfly public key $\overline{K}''$ and the butterfly private key $\overline{K}''^{-1}$ remain a valid key pair. By the same reasoning, it can be shown that, in digital signature applications, the butterfly public key $\overline{K}''$ and the butterfly private key $\overline{K}''^{-1}$ also form a valid pair. For efficient generation of the random invertible matrices $M_{e_1,j}$ and $M_{e_2,j}$, previously published work [16] can be referenced, which describes a method to generate random invertible matrices and their inverses with time complexity O($n$).

$$\begin{aligned} \overline{K}''^{-1}(B) &= \overline{K}''^{-1}\big(\overline{K}''(A)\big), \\ &= \overline{K}'^{-1} \circ M_{e_2,j}{}^{-1}\left(M_{e_2,j} \circ \overline{K}'(A)\right) = \overline{K}'^{-1}\big(\overline{K}'(A)\big), \\ &= \overline{K}^{-1} \circ M_{e_1,j}{}^{-1}\left(M_{e_1,j} \circ \overline{K}(A)\right) = \overline{K}^{-1}\big(\overline{K}(A)\big), \\ &= A = (a_0, a_1, \ldots, a_{n-1}). \end{aligned} \tag{10}$$

*E. Security Discussions*

From the perspective of the RA, the RA knows the relationship between the EE's caterpillar public key $\overline{K}$ and cocoon public key $\overline{K}'$, but does not know the random value $e_2$ or the invertible matrix $M_{e_2,j}$. Therefore, it cannot derive the relationship between the butterfly public key $\overline{K}''$ and the cocoon public key $\overline{K}'$, nor between the butterfly public key $\overline{K}''$ and the caterpillar public key $\overline{K}$. This demonstrates that the EE remains anonymous with respect to the RA.

From the perspective of the CA, the CA knows the relationship between the cocoon public key $\overline{K}'$ and the butterfly public key $\overline{K}''$, but does not know the random value $e_1$ or the invertible matrix $M_{e_1,j}$. Therefore, it cannot derive the relationship between the cocoon public key $\overline{K}'$ and the caterpillar public key $\overline{K}$, nor between the butterfly public key $\overline{K}''$ and the caterpillar public key $\overline{K}$. This demonstrates that the EE remains anonymous with respect to the CA.

From the perspective of other entities, they do not know the random values $e_1$ and $e_2$, nor the invertible matrices $M_{e_1,j}$ and $M_{e_2,j}$. Consequently, they cannot derive the relationship between the butterfly public key $\overline{K}''$ and the cocoon public key $\overline{K}'$, nor between the cocoon public key $\overline{K}'$ and the caterpillar public key $\overline{K}$. This demonstrates that the EE remains anonymous with respect to other entities.

## IV. Experimental Enviorment and Calculation Cases

*A. Experimental Enviorment*

The implementation environment for this study uses the Ubuntu 22.04.3 operating system, with Python 3.10.12 as the programming language and SageMath 9.5 for mathematical computations. Operations such as multivariate calculations, inverse functions, matrix inverses, and finite field arithmetic are all developed and executed based on the SageMath 9.5.

A simple illustrative example is provided in this section; it should be noted that more secure parameter sets should be used in practical deployment. This example is intended solely to demonstrate the feasibility of the method. Let $r = GF(2)$ with cardinality $c = 2$. Let $R$ be a degree-5 extension of $r$, so that $n = 5$. Select the parameter $\tau = 4$, which satisfies the condition $0 < \tau < n$. The elements of the base finite field are $\{0,1\}$. Let $g(x) = x^5 + x^2 + 1$, which is an irreducible polynomial of degree $n$ over $r[x]$. The mapping $K$ is defined as $K(X) = X^{2^4+1} = X^{17}$. Next, by solving the congruence using Eq. (11), the integer $t = 11$ is obtained, yielding the inverse mapping $K^{-1}(X) = X^{11}$.

$$\begin{aligned} t(2^4+1) &\equiv 1\ (mod\ 2^5-1), \\ 17t &\equiv 1\ (mod\ 31), \\ t &\equiv 11\ (mod\ 31). \end{aligned} \tag{11}$$

Sections IV.B through IV.D adopt these parameters and provide computational cases using encryption and decryption as an illustrative application. In these cases, the plaintext is assumed to be encoded as the polynomial $A = \{a_0, a_1, a_2, a_3, a_4\} = \{0,1,0,1,0\} = \begin{bmatrix}1\\0\\0\\0\\1\end{bmatrix}$, which will be used for encryption calculations in the subsequent sections.

*B. Calculational Case 1: Original Key Pair*

As described in Section II, the original public key is defined as $\overline{K}(A) = M_1 \circ \widetilde{K} \circ M_2(A) = M_1 \circ h \circ K \circ h^{-1} \circ M_2(A)$, and encrypting the plaintext $A$ with this public key produces the ciphertext $B$. The original private key is defined as $\overline{K}^{-1}(B) = {M_2}^{-1} \circ \widetilde{K}^{-1} \circ {M_1}^{-1}(B) = {M_2}^{-1} \circ h^{-1} \circ K^{-1} \circ h \circ {M_1}^{-1}(B)$, and decrypting the ciphertext $B$ with this private key recovers the plaintext $A$.

In this section, the matrices are assumed as follows: $M_1 = \begin{bmatrix}0&1&1&0&1\\1&0&0&1&0\\0&0&1&0&1\\1&0&0&0&1\\0&1&0&1&0\end{bmatrix}$ and $M_2 = \begin{bmatrix}0&1&1&0&1\\1&1&0&1&1\\0&0&1&0&0\\1&0&0&0&0\\1&0&0&1&1\end{bmatrix}$. Their inverses are ${M_1}^{-1} = \begin{bmatrix}1&1&1&0&1\\1&0&1&0&0\\1&1&0&1&1\\1&0&1&0&1\\1&1&1&1&1\end{bmatrix}$ and ${M_2}^{-1} = \begin{bmatrix}0&0&0&1&0\\0&1&0&0&1\\0&0&1&0&0\\1&1&1&1&0\\1&1&1&0&1\end{bmatrix}$.

For the encryption calculation, the plaintext $A$ is substituted into the original public key $\overline{K}(A)$. The detailed computational process is shown in Eq. (12) and Eq. (13), resulting in the ciphertext $B = \begin{bmatrix}1\\0\\0\\0\\0\end{bmatrix} = \{1,0,0,0,0\} = \{b_0, b_1, b_2, b_3, b_4\}$.

$$\begin{aligned} \overline{K}(A) &= M_1 \circ \widetilde{K} \circ M_2(A), \\ &= M_1 \circ h \circ K \circ h^{-1} \circ M_2(A), \\ &= M_1 \circ h \circ K \circ h^{-1}\left(\begin{bmatrix}0&1&1&0&1\\1&1&0&1&1\\0&0&1&0&0\\1&0&0&0&0\\1&0&0&1&1\end{bmatrix}\begin{bmatrix}0\\1\\0\\1\\0\end{bmatrix}\right), \\ &= M_1 \circ h \circ K \circ h^{-1}\left(\begin{bmatrix}1\\0\\0\\0\\1\end{bmatrix}\right), \\ &= M_1 \circ h \circ K(1+x^4), \\ &= M_1 \circ h(1+x+x^2+x^3+x^4), \\ &= M_1\left(\begin{bmatrix}1\\1\\1\\1\\1\end{bmatrix}\right) = \begin{bmatrix}0&1&1&0&1\\1&0&0&1&0\\0&0&1&0&1\\1&0&0&0&1\\0&1&0&1&0\end{bmatrix}\begin{bmatrix}1\\1\\1\\1\\1\end{bmatrix} = \begin{bmatrix}1\\0\\0\\0\\0\end{bmatrix}. \end{aligned} \tag{12}$$

$$\begin{aligned} K(1+x^4) &= (1+x^4)^{17}\big(mod\ g(x)\big)(mod\ 2), \\ &\equiv (1+x^4)^{17}\big(mod\ (x^5+x^2+1)\big)(mod\ 2), \\ &\equiv 16625291 - 13925975x + 28289979x^2 \\ &\quad - 23696129x^3 + 19848157x^4 \\ &\quad \big(mod\ g(x)\big)(mod\ 2), \\ &\equiv 1 + x + x^2 + x^3 + x^4. \end{aligned} \tag{13}$$

For the decryption calculation, the ciphertext $B$ is substituted into the original private key $\overline{K}^{-1}(B)$. The detailed computational process is shown in Eq. (14) and Eq. (15), resulting in the plaintext $A = \begin{bmatrix}0\\1\\0\\1\\0\end{bmatrix} = \{0,1,0,1,0\} = \{a_0, a_1, a_2, a_3, a_4\}$, which matches the original plaintext.

$$
\begin{aligned}
\bar{K}^{-1}(B) &= {M_2}^{-1} \circ \tilde{K}^{-1} \circ {M_1}^{-1}(B),\\
&= {M_2}^{-1} \circ h^{-1} \circ K^{-1} \circ h \circ {M_1}^{-1}(B),\\
&= {M_2}^{-1} \circ h^{-1} \circ K^{-1} \circ h\left(\begin{bmatrix}1&1&1&0&1\\1&0&1&0&0\\1&1&0&1&1\\1&0&1&0&1\\1&1&1&1&1\end{bmatrix}\begin{bmatrix}1\\0\\0\\0\\0\end{bmatrix}\right),\\
&= {M_2}^{-1} \circ h^{-1} \circ K^{-1} \circ h\left(\begin{bmatrix}1\\1\\1\\1\\1\end{bmatrix}\right),\\
&= {M_2}^{-1} \circ h^{-1} \circ K^{-1}(1 + x + x^2 + x^3 + x^4),\\
&= {M_2}^{-1} \circ h^{-1}(1 + x^4),\\
&= {M_2}^{-1}\left(\begin{bmatrix}1\\0\\0\\0\\1\end{bmatrix}\right) = \begin{bmatrix}0&0&0&1&0\\0&1&0&0&1\\0&0&1&0&0\\1&1&1&1&0\\1&1&1&0&1\end{bmatrix}\begin{bmatrix}1\\0\\0\\0\\1\end{bmatrix} = \begin{bmatrix}0\\1\\0\\1\\0\end{bmatrix}.
\end{aligned} \tag{14}
$$

$$
\begin{aligned}
&K(1 + x + x^2 + x^3 + x^4),\\
&= (1 + x + x^2 + x^3 + x^4)^{11} \\
&\qquad \big(mod\ g(x)\big)(mod\ 2),\\
&\equiv (1 + x + x^2 + x^3 + x^4)^{17}\\
&\qquad \big(mod\ (x^5 + x^2 + 1)\big)(mod\ 2),\\
&\equiv 22353 + 20668x + 29772x^2 + 12574x^3\\
&\qquad -9147x^4 \big(mod\ g(x)\big)(mod\ 2),\\
&\equiv 1 + x^4.
\end{aligned} \tag{15}
$$

## C. *Calculational Case 2: Expanded Key Pair*

As described in Section III.B, the expanded public key is defined as $\bar{K}'(A) = M_{e_2,j} \circ \bar{K}(A) = M_{e_2,j} \circ M_1 \circ \tilde{K} \circ M_2(A) = M_{e_2,j} \circ M_1 \circ h \circ K \circ h^{-1} \circ M_2(A)$, and encrypting the plaintext $A$ with this public key produces the ciphertext $B'$. The expanded private key is defined as $\bar{K}'^{-1}(B') = \bar{K}^{-1} \circ {M_{e_2,j}}^{-1}(B') = {M_2}^{-1} \circ \tilde{K}^{-1} \circ {M_1}^{-1} \circ {M_{e_2,j}}^{-1}(B') = {M_2}^{-1} \circ h^{-1} \circ K^{-1} \circ h \circ {M_1}^{-1} \circ {M_{e_2,j}}^{-1}(B')$, and decrypting the ciphertext $B'$ with this private key recovers the plaintext $A$.

In this section, the matrix $M_{e_2,j}$ is assumed as $M_{e_2,j} = \begin{bmatrix}0&1&0&0&0\\0&0&0&1&0\\1&0&0&0&0\\0&0&1&0&0\\0&0&0&0&1\end{bmatrix}$, and the inversed matrix ${M_{e_2,j}}^{-1}$ is ${M_{e_2,j}}^{-1} = \begin{bmatrix}0&0&1&0&0\\1&0&0&0&0\\0&0&0&1&0\\0&1&0&0&0\\0&0&0&0&1\end{bmatrix}$.

For the encryption calculation, the plaintext $A$ is substituted into the expanded public key $\bar{K}'(A)$. The detailed computational process is shown in Eq. (16), resulting in the ciphertext $B' = \begin{bmatrix}0\\0\\1\\0\\0\end{bmatrix} = \{0,0,1,0,0\} = \{b_0, b_1, b_2, b_3, b_4\}$.

$$
\begin{aligned}
\bar{K}'^{(A)} &= M_{e_2,j} \circ \bar{K}(A),\\
&= M_{e_2,j}\left(\begin{bmatrix}1\\0\\0\\0\\0\end{bmatrix}\right) = \begin{bmatrix}0&1&0&0&0\\0&0&0&1&0\\1&0&0&0&0\\0&0&1&0&0\\0&0&0&0&1\end{bmatrix}\begin{bmatrix}1\\0\\0\\0\\0\end{bmatrix} = \begin{bmatrix}0\\0\\1\\0\\0\end{bmatrix}.
\end{aligned} \tag{16}
$$

For the decryption calculation, the ciphertext $B'$ is substituted into the expanded private key $\bar{K}'^{-1}(B')$. The detailed computational process is shown in Eq. (17), resulting in the plaintext $A = \begin{bmatrix}0\\1\\0\\1\\0\end{bmatrix} = \{0,1,0,1,0\} = \{a_0, a_1, a_2, a_3, a_4\}$, which matches the original plaintext.

$$
\begin{aligned}
\bar{K}'^{-1}(B') &= \bar{K}^{-1} \circ {M_{e_2,j}}^{-1}(B'),\\
&\bar{K}^{-1}\left(\begin{bmatrix}0&0&1&0&0\\1&0&0&0&0\\0&0&0&1&0\\0&1&0&0&0\\0&0&0&0&1\end{bmatrix}\begin{bmatrix}0\\0\\1\\0\\0\end{bmatrix}\right),\\
&= {M_2}^{-1} \circ h^{-1} \circ K^{-1} \circ h \circ {M_1}^{-1}\left(\begin{bmatrix}1\\0\\0\\0\\0\end{bmatrix}\right) = \begin{bmatrix}0\\1\\0\\1\\0\end{bmatrix}.
\end{aligned} \tag{17}
$$

## D. *Calculational Case 3: Butterfly Key Pair*

As described in Section III.C, the butterfly public key is defined as $\bar{K}''(A) = M_{e_2,j} \circ M_{e_1,j} \circ \bar{K}(A) = M_{e_2,j} \circ M_{e_1,j} \circ M_1 \circ \tilde{K} \circ M_2(A) = M_{e_2,j} \circ M_{e_1,j} \circ M_1 \circ h \circ K \circ h^{-1} \circ M_2(A)$, and encrypting the plaintext $A$ with this public key produces the ciphertext $B''$. The butterfly private key is defined as $\bar{K}''^{-1}(B'') = \bar{K}^{-1} \circ {M_{e_1,j}}^{-1} \circ {M_{e_2,j}}^{-1}(B'') = {M_2}^{-1} \circ \tilde{K}^{-1} \circ {M_1}^{-1} \circ {M_{e_1,j}}^{-1} \circ {M_{e_2,j}}^{-1}(B'') = {M_2}^{-1} \circ h^{-1} \circ K^{-1} \circ h \circ {M_1}^{-1} \circ {M_{e_1,j}}^{-1} \circ {M_{e_2,j}}^{-1}(B'')$, and decrypting the ciphertext $B''$ with this private key recovers the plaintext $A$.

In this section, the matrix $M_{e_1,j}$ is assumed as $M_{e_1,j} = \begin{bmatrix}0&0&0&1&0\\0&1&0&0&0\\0&0&0&0&1\\1&0&0&0&0\\0&0&1&0&0\end{bmatrix}$, and the inversed matrix ${M_{e_1,j}}^{-1}$ is ${M_{e_1,j}}^{-1} = \begin{bmatrix}0&0&0&1&0\\0&1&0&0&0\\0&0&0&0&1\\1&0&0&0&0\\0&0&1&0&0\end{bmatrix}$.

For the encryption calculation, the plaintext $A$ is substituted into the butterfly public key $\bar{K}'(A)$. The detailed computational process is shown in Eq. (18), resulting in the ciphertext $B'' = \begin{bmatrix}0\\1\\0\\0\\0\end{bmatrix} = \{0,1,0,0,0\} = \{b_0, b_1, b_2, b_3, b_4\}$.

$$\begin{aligned}
\bar{K}''(A) &= M_{e_2,j} \circ M_{e_1,j} \circ \bar{K}(A),\\
&= M_{e_2,j} \circ M_{e_1,j}\left(\begin{bmatrix}1\\0\\0\\0\\0\end{bmatrix}\right),\\
&= M_{e_2,j}\left(\begin{bmatrix}0&0&0&1&0\\0&1&0&0&0\\0&0&0&0&1\\1&0&0&0&0\\0&0&1&0&0\end{bmatrix}\begin{bmatrix}1\\0\\0\\0\\0\end{bmatrix}\right),\\
&= M_{e_2,j}\left(\begin{bmatrix}0\\0\\0\\1\\0\end{bmatrix}\right) = \begin{bmatrix}0&1&0&0&0\\0&0&0&1&0\\1&0&0&0&0\\0&0&1&0&0\\0&0&0&0&1\end{bmatrix}\begin{bmatrix}0\\0\\0\\1\\0\end{bmatrix} = \begin{bmatrix}0\\1\\0\\0\\0\end{bmatrix}.
\end{aligned} \tag{18}$$

For the decryption calculation, the ciphertext $B''$ is substituted into the butterfly private key $\bar{K}''^{-1}(B'')$. The detailed computational process is shown in Eq. (19), resulting in the plaintext $A = \begin{bmatrix}0\\1\\0\\1\\0\end{bmatrix} = \{0,1,0,1,0\} = \{a_0, a_1, a_2, a_3, a_4\}$, which matches the original plaintext.

$$\begin{aligned}
\bar{K}''^{-1}(B'') &= \bar{K}^{-1} \circ M_{e_1,j}{}^{-1} \circ M_{e_2,j}{}^{-1}(B''),\\
&= \bar{K}^{-1} \circ M_{e_1,j}{}^{-1} \circ M_{e_2,j}{}^{-1}\left(\begin{bmatrix}0\\1\\0\\0\\0\end{bmatrix}\right),\\
&= \bar{K}^{-1} \circ M_{e_1,j}{}^{-1}\left(\begin{bmatrix}0&0&1&0&0\\1&0&0&0&0\\0&0&0&1&0\\0&1&0&0&0\\0&0&0&0&1\end{bmatrix}\begin{bmatrix}0\\1\\0\\0\\0\end{bmatrix}\right),\\
&= \bar{K}^{-1}\left(\begin{bmatrix}0&0&0&1&0\\0&1&0&0&0\\0&0&0&0&1\\1&0&0&0&0\\0&0&1&0&0\end{bmatrix}\begin{bmatrix}0\\0\\0\\1\\0\end{bmatrix}\right),\\
&= M_2{}^{-1} \circ h^{-1} \circ K^{-1} \circ h \circ M_1{}^{-1}\left(\begin{bmatrix}1\\0\\0\\0\\0\end{bmatrix}\right) = \begin{bmatrix}0\\1\\0\\1\\0\end{bmatrix}.
\end{aligned} \tag{19}$$

## V. Conclusions

This study proposes a multivariate cryptography-based anonymous certificate scheme. Building upon the principles of multivariate cryptography, a key expansion method and an anonymous certificate scheme are designed, and the butterfly key expansion is incorporated to achieve anonymity with respect to both the RA and the CA. Mathematical proofs are provided, along with computational examples to demonstrate the feasibility of the proposed methods. The results indicate that the proposed approach can achieve anonymity without increasing ciphertext or signature lengths.